# Ballistic molecular transport through two-dimensional channels

A. Keerthi[1,2], A. K. Geim[1,2]*, A. Janardanan[1], A. P. Rooney[3], A. Esfandiar[2,4], S. Hu[2], S. A. Dar[2],
I. V. Grigorieva[1], S. J. Haigh[3], F. C. Wang[1,5], B. Radha[1,2]*

[1]School of Physics & Astronomy, University of Manchester, Manchester M13 9PL, UK
[2]National Graphene Institute, University of Manchester, Manchester M13 9PL, UK
[3]School of Materials, University of Manchester, Manchester, M13 9PL, UK
[4]Department of Physics, Sharif University of Technology, P.O. Box 11155-9161, Tehran, Iran
[5]Chinese Academy of Sciences Key Laboratory of Mechanical Behavior & Design of Materials, Department of Modern Mechanics, University of Science & Technology of China, Hefei 230027, China

**Gas permeation through nanoscale pores is ubiquitous in nature and plays an important role in a plethora of technologies[1,2]. Because the pore size is typically smaller than the mean free path of gas molecules, their flow is conventionally described by the Knudsen theory that assumes diffuse reflection (random-angle scattering) at confining walls[3-7]. This assumption has proven to hold surprisingly well in experiment, and only a few cases of partially specular (mirror-like) reflection are known[5,8-11]. Here we report gas transport through angstrom-scale channels with atomically-flat walls[12,13] and show that surface scattering can be both diffuse or specular, depending on fine details of the surface atomic landscape, and quantum effects contribute to the specularity at room temperature. The channels made from graphene or boron nitride allow a helium gas flow that is orders of magnitude faster than expected from the theory. This is explained by specular surface scattering, which leads to ballistic transport and frictionless gas flow. Similar channels but with molybdenum disulfide walls exhibit much slower permeation that remains well described by Knudsen diffusion. The difference is attributed to stronger atomic corrugations at $MoS_2$ surfaces, which are similar in height to the size of transported atoms and their de Broglie wavelength. The importance of the latter, matter-wave contribution is corroborated by the observation of a reversed isotope effect in which the mass flow of hydrogen is notably higher than that of deuterium, in contrast to the relation expected for classical flows. Our results provide insights into atomistic details of molecular permeation, which so far could be accessed only in simulations[10,14], and show a possibility of studying gas transport under a controlled confinement comparable to the quantum-mechanical size of atoms.**

The Knudsen theory provides comprehensive description of gas flow in the regime where molecules collide mostly with confining walls rather than each other. Despite its universal adoption, the theory relies on certain assumptions including fully diffusive surface scattering[3,5,7]. Recently, several new experimental systems with nanoscale channels have been introduced, including carbon nanotubes[8,11] and nanoporous films made from graphene[15-19], graphene oxide[19,20] and other two-dimensional (2D) materials[21,22]. A number of anomalies in the gas permeation properties were reported, which in certain cases[8,11] were difficult to reconcile within the classical theory. In particular, the observation of fast gas flows through narrow carbon nanotubes was attributed to a combination of specular and diffusive scattering[8,11]. Unfortunately, the exact dimensions and structure of these new systems are often insufficiently controlled, which makes it difficult to compare the observed behavior with a large and growing body of molecular dynamics simulations[10,14]. The analysis is further complicated by poorly understood effects of nanotubes' curvature[23] and, especially, the presence of adsorbates (hydrocarbons, etc.) which universally cover surfaces that are not under ultrahigh vacuum[9,24-26].

In this work, we report gas transport through angstrom-scale slit-like channels with walls made from cleaved crystals of graphite, hexagonal boron nitride (hBN) or molybdenum disulfide ($MoS_2$). These three materials were chosen as archetypal examples of crystals that can be exfoliated down to monolayers and provide atomically-flat surfaces stable under ambient conditions[12]. The nanochannels were fabricated following the recipe described in Supplementary Information section 1. In brief, two thin (~10-100 nm) crystals of the above materials were prepared by exfoliation to serve as bottom and top walls of an intended channel. The third, thinner crystal was plasma-etched to contain long narrow trenches (Fig. 1). It served as a spacer between the top and bottom walls. The three crystals were assembled on top of each other as shown in Fig. 1a, being held together by van der Waals (vdW) forces[12]. The slit height $h$ was determined by the vdW thickness of the spacer crystal and could be chosen from being just one atomic layer up to as many as required. In all the measurements reported below, the trench width $w$ defining the width of the resulting channels was ~130 nm, chosen to be sufficiently large to allow accurate measurements of a gas flow but not so large as to allow sagging[13]. To increase the measurement accuracy, we often used many slits in parallel (typically, 200) but some experiments were also done using individual channels. Their length $L$ was defined by lithography from ~1 to > 15 μm (Supplementary Fig. 1). An example of our 2D slits is shown in Figs. 1a,b where $MoS_2$ is intentionally used as the building material to provide high contrast for transmission electron microscopy (TEM) imaging. One can view this channel as if a single atomic plane is removed from a bulk $MoS_2$ crystal, which results in a pair of edge dislocations and a 2D empty space with $h \approx 6.7$ Å between them. Slits using different wall materials were made and studied (Supplementary Information), and we mainly used graphene spacers that allowed $h = N \times a$ in multiples of the graphene vdW thickness $a \approx 3.4$ Å, where $N$ is the number of layers.

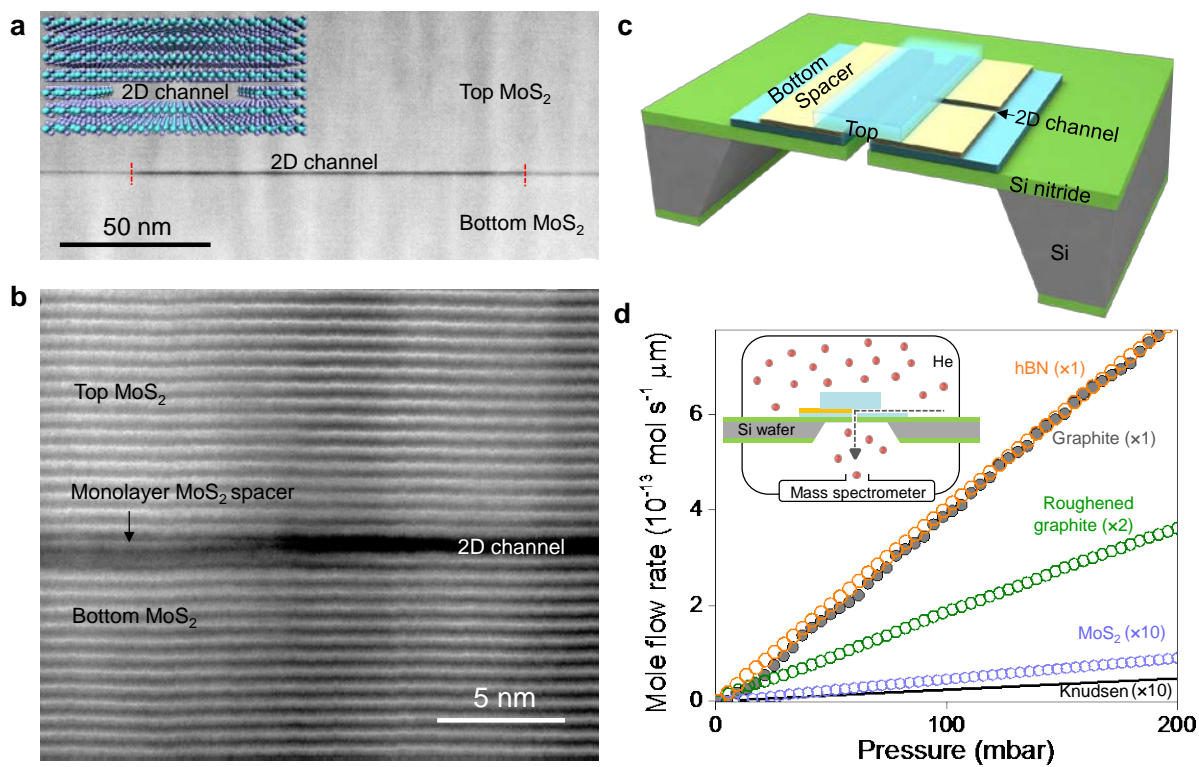

**Figure 1 | Angstrom-scale slits and helium gas transport through them. a,** 2D channel assembled from $MoS_2$ crystals. Inset and main panel: Its schematic and TEM micrograph, respectively. The channel is seen in black. For clarity, its edges are marked with red ticks. The monolayer spacer appears somewhat darker with respect to the top and bottom crystals because of different in-plane orientations. The contrast ripples running vertically are the curtaining effect that occurs during ion beam polishing[27]. Micrographs of slits made from other 2D materials can be found in Supplementary Fig. 2 and ref. 13. **b,** High magnification image of the channel in **a** near its left edge. Each bright horizontal line corresponds to monolayer $MoS_2$. **c,** Schematic of our experimental devices. The tri-crystal assembly (cyan and yellow) covers an aperture in a silicon nitride membrane (green) prepared on top of a Si wafer (grey). **d,** Comparison of He permeation through 2D channels of the same height ($N$ = 5) but with walls made from different crystals. All the devices here are single-channel with $L$ ranging from 1 to 6 μm. The mole flow rates at room temperature (296±3K) are normalized per channel length and, for legibility, multiplied by the factors shown in the color-coded brackets. The flow expected for Knudsen diffusion is shown by the solid line close to the $MoS_2$ data. Inset: Our measurement setup. The arrow indicates the gas flow direction.

For measurements of gas transport through the described 2D channels, we made devices shown schematically in Fig. 1c. The vdW assembly was placed to seal a micrometer-size opening in a silicon wafer (Supplementary Figs 1 and 3). The wafer separated two containers, one of which had a gas under the variable pressure $P$ whereas the other was a vacuum chamber equipped with a mass spectrometer. Unless specifically stated below, we used helium as a test gas. The applied pressure $P$ was slowly increased up to 200 mbar, which corresponded to the mean free path always larger than 0.7 μm and typical Knudsen numbers > $10^4$. Examples of the measured flow rates $Q$ at room temperature as a function of $P$ are shown in Fig. 1d. As a reference, we made devices using the same fabrication procedures but without etching trenches in the spacer crystal. Those devices exhibited no discernable He permeation, confirming that the 2D channels were the only pathway between the two containers. The minimal spacers that allowed us to detect gas transport were found to be monolayer MoS$_2$ ($h \approx 6.7$ Å) as in Fig. 1 and bilayer graphene ($N = 2$; $h \approx 6.8$ Å).

Helium transport through our 2D channels was found to depend strongly on their wall material. Indeed, Fig. 1d shows that devices of exactly the same geometry ($h = 5a \approx 17$ Å) exhibited $Q$ that were two orders of magnitude larger for graphene and hBN walls than for MoS$_2$. In our case where the mean free path of He atoms is much larger than $h$, the Knudsen theory expects the mass flow rate[5,6]

$$Q_K = \alpha P \, (m/2\pi RT)^{1/2} wh \quad (1)$$

where $m$ is the atomic mass of the transported gas, $R$ the gas constant, $T$ the temperature and $\alpha$ the transmission coefficient. For long and narrow rectangular channels ($h << w < L$), $\alpha$ can be approximated as[6]

$$\alpha \approx (h/L) \ln(4w/h) \quad (2).$$

According to these equations, the $N = 5$ channels in Fig. 1d should have exhibited $Q$ shown by the solid line. The Knudsen prediction holds (within a factor of 2) only for MoS$_2$ walls. The other 2D slits exhibited much higher $Q$. Similar disparity was observed for other $h$ (see below). Note that the large aspect ratios $w/h \approx 100$ imply that gas molecules collide mostly with top and bottom walls and scattering at sidewalls plays a relatively little role, in agreement with our experimental observation that channels made using different sidewall materials but similar $h$ exhibited close $Q$ (see, e.g., channels in Fig. 1 and Supplementary Fig. 2b).

To quantify the observed enhancement with respect to the Knudsen theory, we introduce the enhancement coefficient $K = Q/Q_K$. This allows us to summarize our finding in Fig. 2a where results for more than 70 devices are plotted. The figure shows that within the data scatter the gas flow through MoS$_2$ channels is described well by eqs (1-2). In contrast, all graphite and hBN devices exhibited strong enhancement reaching $K > 100$. There is a clear tendency to smaller $K$ with increasing $N$ such that $K$ decreases to ~10 and ~3 for $N = 12$ and 25, respectively. This is perhaps not surprising because large channels are eventually expected to follow the standard behavior. Let us emphasize that the enhancement effect is so strong that graphene and hBN slits with $h \approx 1.4$ nm exhibited a He flux ~ 10 times larger than the 9 nm slits, in contrast to eqs (1-2) which yield a difference by a factor of ~ 10 but in the opposite direction.

The observed flow enhancement can be attributed to specular reflection of He atoms off atomically-flat walls, which should result in the breakdown of Knudsen's approximation. This possibility was first evoked by Smoluchowski[4] who modified the theory by introducing the fraction $f$ of diffusely reflected molecules. In this case, the transmission coefficient $\alpha$ should - in the first approximation[4,28] - be multiplied by the factor $(2-f)/f$. Using the Smoluchowski model[4,5,10] our data yield, for example, $f \approx 0.2$ for graphite and hBN channels with $h \approx$ 4 nm and $f$ of about 1 for all MoS$_2$ devices. The largest $K$ in Fig. 2a suggest $f$ very close to zero, that is, specular reflection. Moreover, our data for graphene channels with $N = 4$ yield practically perfect transmission without invoking the Smoluchowski theory. This limit ($\alpha = 1$) is plotted in Fig. 2a and corresponds to a gas flow through an aperture of size $w \times h$ in an infinitely thin membrane. The observed agreement means that helium atoms pass through our long slits ballistically, without losing their momenta.

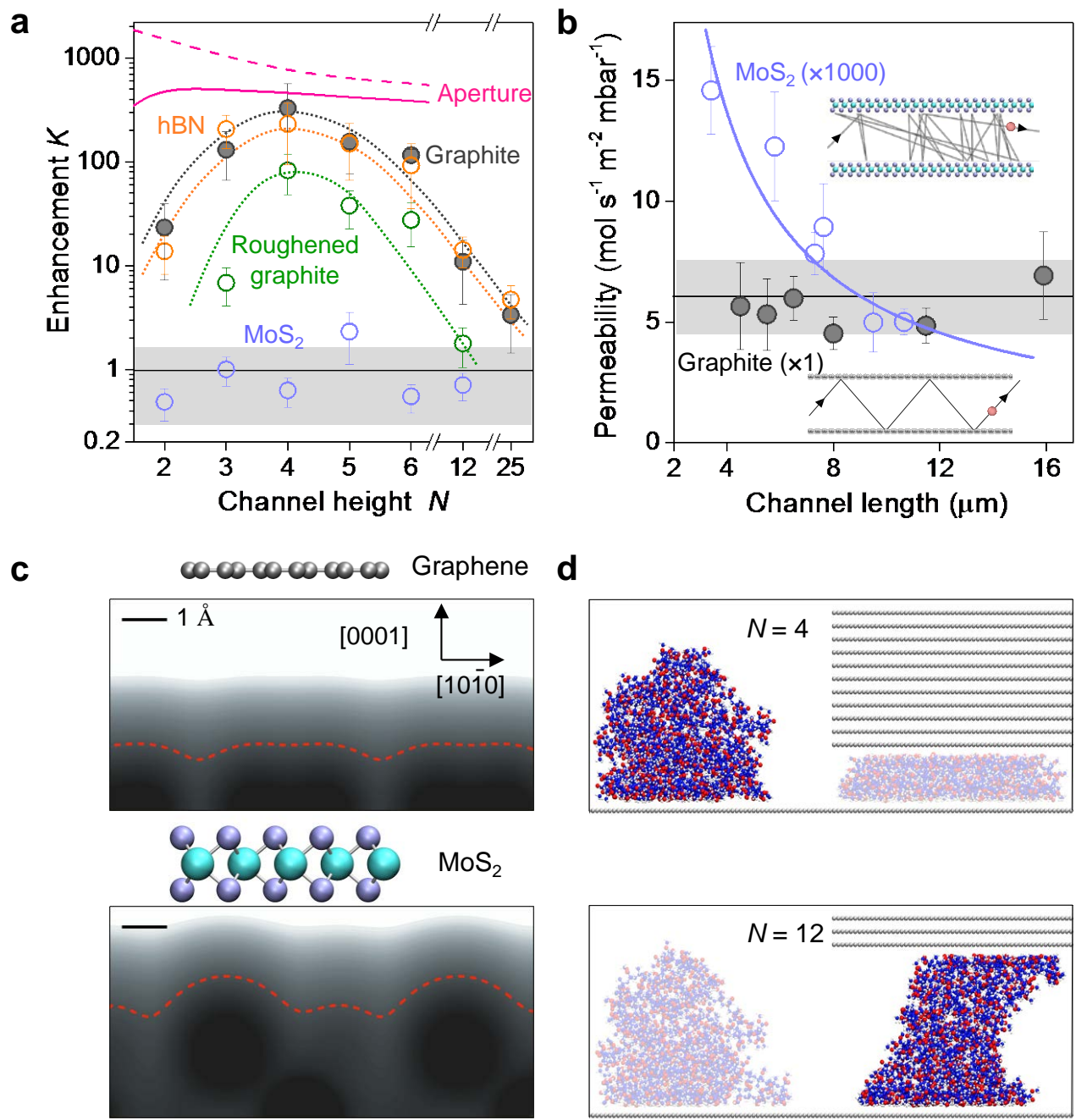

**Figure 2 | Enhanced gas flow through 2D channels. a,** $K$ observed for different walls and heights (symbols). Those could have different $L$. Dotted curves: Guides to the eye. MoS$_2$ channels exhibit no enhancement within our data scatter. Dashed curve: $K$ expected from eq. (1) for ideal transmission. The solid curve is also for $\alpha = 1$ but takes into account the finite size $d$ of He atoms (Supplementary Information section 6). **b,** Dependence of He transport on $L$; $N = 4$. In the case of MoS$_2$ walls, the flow follows the $1/L$ dependence expected for diffusive transport (purple curve; the data are multiplied by a factor of 1,000). No length dependence is found for graphene channels, a clear signature of frictionless transport. The insets illustrate diffusive and ballistic scattering. Error bars in **a**, **b**: Statistical errors from our measurements using at least two devices for each $N$. The shaded areas in **a**, **b** indicate the standard error for the corresponding data sets. **c,** Intrinsic roughness of atomically-flat surfaces. Grey scale: Electron density near graphene and MoS$_2$ surfaces. Red curves: Depth accessible for He atoms with thermal energies (Supplementary Information section 4). **d,** Illustration of favorable and unfavorable positions (bright and faint, respectively) for a polymer molecule inside slits with different $N$.

To further substantiate the observed frictionless flow, we carried out two additional experiments. First, we made graphene channels with 'roughened' bottom walls using a short exposure to oxygen plasma (Supplementary Information section 1). The induced roughness suppressed the He flow by an order of magnitude, albeit insufficient to recover the Knudsen description (Figs. 1d and 2a). Non-zero specularity can be attributed to the fact that the oxygen plasma tends to etch holes in graphite so that a large portion of its surface remains flat[29]. The clearest evidence for ballistic transport comes however from the second set of experiments in which we used channels of the same height $h = 4a$ (highest $K$) but with different $L$. Fig. 2b shows that the gas flow through the channels with $MoS_2$ walls decays with increasing $L$ and exhibits the $Q \propto 1/L$ dependence as prescribed by the standard theory. In stark contrast, graphene slits flaunt the gas flow independent of $L$, as expected for ballistic transport.

The profound difference between the devices with $MoS_2$ and graphene/hBN walls is surprising. Indeed, cleaved $MoS_2$ may not be as perfect as graphene but its surface still contains few defects[28], especially in comparison with roughened graphene. We believe that the difference arises from a finite roughness of ideal, atomically-flat surfaces. Fig. 2c shows that $MoS_2$ exhibits relatively strong corrugations, reaching ∼ 1 Å in height, such that incoming He atoms should be able to 'see' this roughness because its scale is comparable to both helium's kinetic diameter, $d \approx 2.6$ Å, and its de Broglie wavelength, $\lambda_B \approx 0.5$ Å at room temperature. The former is a semiclassical notion but, nonetheless, represents the quantum-mechanical size of the electron cloud around helium nuclei[31]. Graphene and hBN surfaces are much flatter on this scale (Fig. 2c and Supplementary Fig. 4), suggesting more specular reflection.

The above experiments cannot distinguish between the effects of $d$ and $\lambda_B$ and, to figure out whether quantum effects contribute to the observed specular reflection, we measured the isotope effect by comparing hydrogen and deuterium permeation. The isotopes have the same $d$ and the same interaction with the walls but different $\lambda_B$. Eq. (1) suggests that $Q$ should be $\sqrt{2}$ larger for deuterium, independent of experimental details (channel geometry, $f$, etc.) This benchmark dependence was validated using micrometer-size apertures (Supplementary Information section 3). In contrast, our graphene channels with $N = 4$ exhibited $Q \approx 30\pm10\%$ smaller for deuterium than hydrogen. This fact unequivocally proves that matter-wave effects contribute to the specular reflection leading to its suppression for heavier atoms because deuterium has a shorter $\lambda_B$ and sees an atomic landscape somewhat rougher than hydrogen.

Finally, let us discuss the changeover from ballistic to classical (diffusive) transport, which is observed for relatively large $h \approx$ 4-8 nm (Fig. 2a). We believe that the underlying reason is that such large channels are no longer atomically flat due to hydrocarbon contamination. Indeed, cleaved surfaces are rapidly covered with various adsorbates even under cleanroom conditions[24-26]. This is particularly obvious in a high-resolution TEM where only tiny areas of graphene can be found devoid of hydrocarbons[24]. However, if two atomically-flat surfaces are brought together, this contamination is known to aggregate into well separated pockets (so-called contamination bubbles) such that outside them the attached surfaces become atomically clean, free of any adsorbates[12,25]. We believe that this self-cleansing occurs already at a finite separation between 2D crystals. Polymer molecules tend to arrange themselves into clumps of a few nm in height (Fig. 2d and Supplementary Information section 5). Confining such clumps between parallel walls reduces their configurational entropy that competes with an energy gain due to adhesion (Supplementary Fig. 5). The latter is a surface effect whereas the former is determined by the clumps' volume, which implies that the squashed clumps become energetically unfavorable at small $h$ and should be squeezed out. Our molecular dynamics simulations (Supplementary Information section 5) show that, polymer molecules prefer to sit outside narrow channels whereas their interior position is more favorable for larger $h$ (Fig. 2d). As for deviations from ballistic transport for our thinnest channels ($N = 2$), we speculate that the reduced transmission can be due to another size effect. Indeed, the available empty space not filled with electron clouds is ∼ 6.7 Å, only slightly larger than $d \approx 2.6$ Å (Supplementary Fig. 6). In combination with a finite $\lambda_B \approx 0.5$ Å that effectively increases $d$, this should decrease the number of atoms that can enter the narrow aperture, especially at shallow incident angles, as discussed in Supplementary Information section 6 (solid curve in Fig. 2a). Interplay between the two mechanisms that suppress molecular flow at both large and small $N$ leads to the observed dome-shaped dependence. Further theory and simulations are needed to analyze such size effects in detail.

To conclude, our work offers new understanding for many previous predictions, calculations and observations. For example, it seems to reconcile widely-varying results for molecular transport through carbon nanotubes. A strongly enhanced gas flow[8,11] was reported for sub-2 nm nanotubes whereas no such enhancement was found by other groups for wider tubes, in conceptual agreement with the observed transition from ballistic to diffusive transport due to the increasing role of hydrocarbon contamination in larger channels. It is also instructive to point out an analogy with ballistic electron transport in metallic systems. They can exhibit a finite electrical conductance even in the absence of electron scattering, which is known as the point contact resistance[32]. The effect is described by the Sharvin formula[32] that is equivalent to eq. (1) with $\alpha \equiv 1$ but for a charge rather than gas flow. Ballistic effects in molecular transport have not been considered previously even in theory but the demonstrated angstrom-scale channels now make this regime accessible experimentally.

# Supplementary

**1. Making 2D channels**

Our devices were made using the series of microfabrication procedures, which is illustrated in Supplementary Fig. 1. First, a free-standing silicon nitride ($SiN_x$) membrane with dimensions of ~ 100 μm × 100 μm was prepared starting with the standard Si wafer covered with a 500 nm thick layer of $SiN_x$. This was done using photolithography and wet etching. A rectangular hole of ~ 3 μm × 26 μm in size was then made in this membrane using photolithography and dry etching. Next a thin (~10-30 nm) crystal of graphite, hBN or $MoS_2$ was mechanically exfoliated and transferred[12,13] onto the membrane to cover the opening (Supplementary Fig. 1a). This crystal served as the bottom layer in our trilayer-crystal assembly. Following the transfer, the rectangular hole was extended into the bottom layer using the $SiN_x$ membrane as a mask for dry etching from the back side of the Si wafer (Supplementary Fig. 1). To this end, oxygen plasma was employed for etching graphite whereas hBN and $MoS_2$ were etched in a mixture of $CHF_3$ and oxygen.

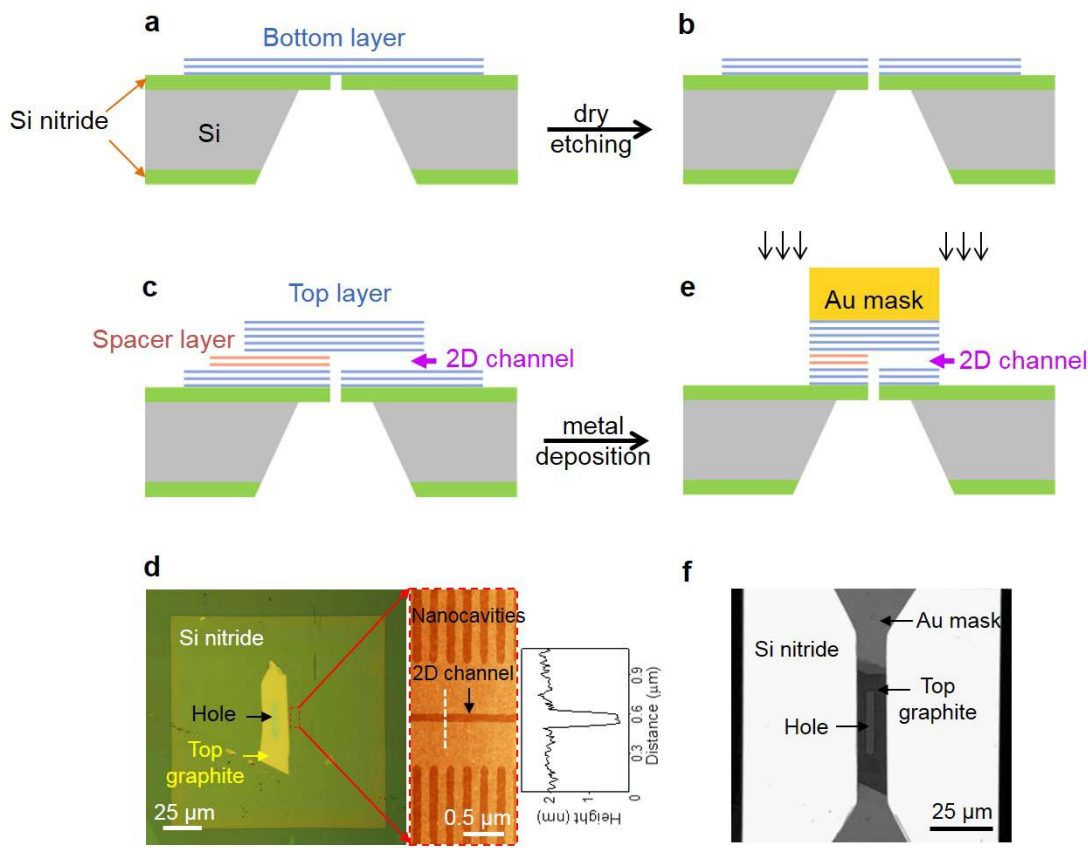

**Supplementary Figure 1| Fabrication procedures. a,** Thin crystal was transferred to cover an opening in a $SiN_x$ membrane. **b,** The opening is extended through the bottom crystal. **c,** Spacer stripes were deposited on top of the bottom layer and usually etched from the back side. The top crystal is then transferred on top to fully cover the rectangular opening. **d,** Left: Optical image of a single-channel capillary device made entirely from graphite; $N$ = 5. The $SiN_x$/Si wafer is seen in dark green; the $SiN_x$ membrane appears as a light-green square; the top graphite layer shows in bright yellow and the rectangular opening (lighter green) is indicated by the black arrow. Center: Atomic force micrograph near the channel entry where the top graphite does not cover the spacer layer (scan area is shown by the red contour). The height profile taken along the dotted white line is shown to the right, indicating the channel height $h \approx 1.7 \pm 0.1$ nm. The side cavities perpendicular to the 2D channel were made to prevent contamination bubbles[25] across the main channel. **e,** A gold mask is placed on top of the trilayer assembly for final etching to define the length $L$ and unblock the channel entry. **f,** Optical image of the final capillary device in the transmission mode. The $SiN_x$ membrane is fully transparent (bright). The Au mask is partially transparent, and both top graphite and the rectangular hole in $SiN_x$ can be seen underneath Au as indicated by the arrows.

To make the second (spacer) layer, 2D crystals of graphene or $MoS_2$ were exfoliated onto an oxidized Si wafer (300 nm of $SiO_2$). Crystals of a chosen thickness were then etched into stripes of ~130 nm in width and separated by the same distance. This was done using electron beam lithography (polymethyl methacrylate (PMMA) with molecular weight 950K was employed as a resist) and plasma etching. The PMMA mask was removed by mild sonication in acetone. The resulting stripes were transferred on top of the bottom layer as shown in Supplementary Fig. 1c,d. Next, a relatively thick (~100 nm) crystal of graphite, hBN or $MoS_2$ was dry-transferred[13,33] on top of the two-layer assembly so that it covered the rectangular hole and partially overlapped with the spacer stripes. After each layer transfer we annealed our assembly in 10%-hydrogen-in-argon at 300-400°C for 3 hours. The annealing step was critical for cleanliness of the final devices to avoid channels' blockage with a PMMA residue.

For the experiment that used roughened channels, the bottom graphite crystal was briefly exposed to oxygen plasma to remove approximately 3 layers of graphene. This was done before transferring the spacer layer. To define the length $L$ of the final channels, a metal mask (5 nm Cr/50 nm Au) was placed by photolithography on top of the final assembly as shown in Supplementary Fig. 1e,f. Dry etching through this mask not only allowed us to control $L$ accurately but also opened channels' entries if those were accidentally blocked by thin (<10 nm) edges of the top crystal, which tended to sag inside the channels[13]. For single-channel devices used in some experiments (e.g., see Fig. 1d of the main text), extra nanocavities (Supplementary Fig. 1d) were created around the main channel to prevent the formation of bubbles that collected hydrocarbons and other contaminants[12,25,33] and could block individual channels. These cavities were arranged perpendicular to the main channel (Supplementary Fig. 1d) and did not contribute into gas transport through the final devices.

**2. Cross-sectional imaging of 2D channels**

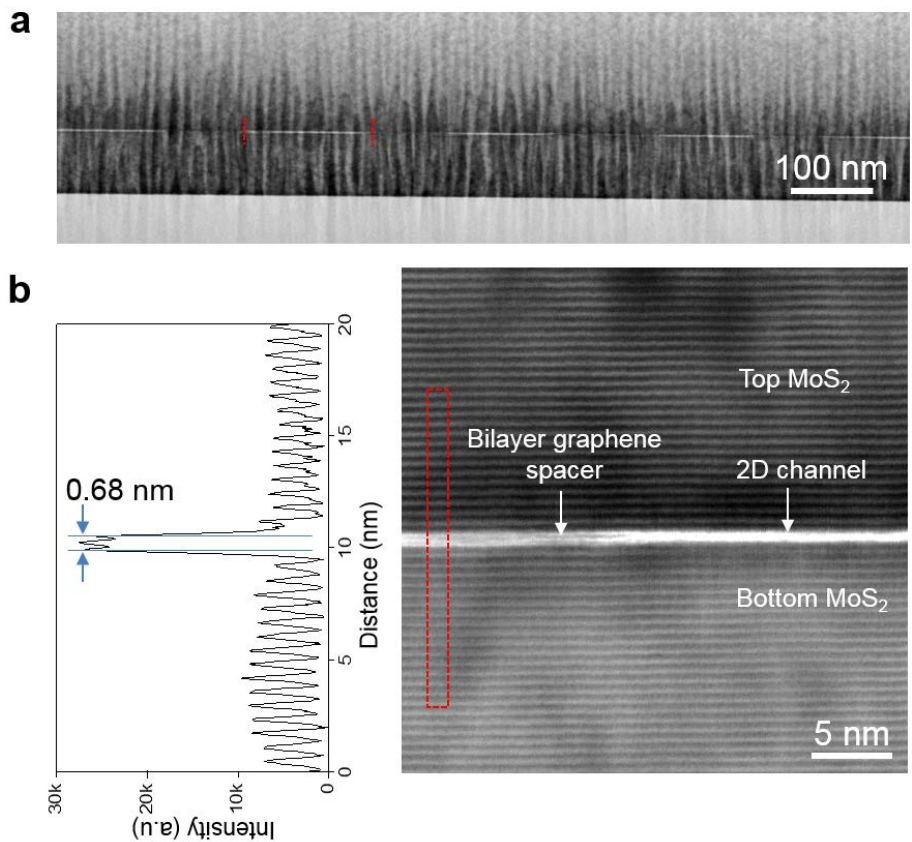

**Supplementary Figure 2| Visualization of 2D channels. a,** Array of 2D slits made entirely from $MoS_2$ as imaged in STEM bright-field. For guidance, edges of one of the channels are indicated by red marks. The vertical stripes are the known curtaining effect causes by ion milling[27]. **b,** High magnification STEM image of a 2D channel with the top and bottom walls made from $MoS_2$ and bilayer graphene as the spacer (right image). The channel is white in the bright-field image, and atomic layers of $MoS_2$ can be seen as dark lines running parallel to it. The left panel shows a contrast profile across the region indicated by the red rectangle. Cross-sectional images of 2D slits made from graphite crystals can be found in ref. 13.

For scanning transmission electron microscopy (STEM) and high angle annular dark field (HAADF) imaging, we made thin cross-sectional lamellae implementing an in situ lift-out procedure[25,34]. Lamellae were cut out perpendicular to the capillary axis by high-precision site-specific milling in Helios Nanolab DualBeam 660 that incorporates scanning electron microscope and focused ion beam columns. Platinum was deposited using the ion beam to weld the lamella to a micromanipulator, which is then lifted from the substrate. Once transferred to a specialist OmniProbe TEM grid, the lamella foil was thinned down to < 100 nm and then polished further to electron transparency, using 5 kV and subsequently 2 kV ion milling. High-resolution STEM and HAADF images were acquired in an aberration-corrected microscope (FEI Titan G2 80-200 kV) using a probe convergence angle of 21 mrad, a HAADF inner angle of 48 mrad and a probe current of ~70 pA. To ensure that the electron beam was parallel to the 2D channels, it was aligned using the relevant Kikuchi bands of the silicon substrate and the assembled 2D crystals.

**3. Gas transport measurements**

Gas permeation through our 2D slits was studied in the steady state flow regime using the setup shown in Supplementary Fig. 3a. The $SiN_x$/Si wafer containing a capillary device was sealed using O-rings to separate two oil-free vacuum chambers. They were evacuated before every experimental run. One of the chambers was equipped with an electrically-controlled dosing valve that provided the pressure $P$ inside, which was monitored by a pressure gauge. The top (entry) side of our devices was facing this chamber (Supplementary Fig. 3a). The devices were sufficiently robust to withstand $P$ up to 2 bars. The other chamber was maintained at a pressure of $\sim 10^{-6}$ bar, and connected to a mass spectrometer. For flow measurements of helium, hydrogen and deuterium, we used a calibrated helium-leak detector (INFICON UL200) as the mass spectrometer. The leak detector measures the flow rates in mbar L $s^{-1}$, which are straightforward to convert into mol $s^{-1}$ using the ideal gas equation. All the measurements were done at room temperature ($T$ = 296 ± 3K) as measured by a probe mounted close to the device.

The measurement setup was thoroughly checked for possible leaks using control devices that were prepared following the same fabrication procedures but contained no 2D slits ($N$ = 0). They exhibited no discernible He leak, which proved that 2D channels were the only possible permeation path for the tested gases. To assure quantitative accuracy of our measurements, we also prepared reference devices containing round apertures made in $SiN_x$ membranes and tested their conductance with respect to helium. The measured flow values were essentially indistinguishable from those given by the Knudsen equation (Supplementary Fig. 3b). For other gases, the apertures were used to calibrate the sensitivity of our mass spectrometer and, also, to measure the isotope effect in which the mass flow of deuterium was observed to be √2 higher than that of hydrogen, as required by eq. (1).

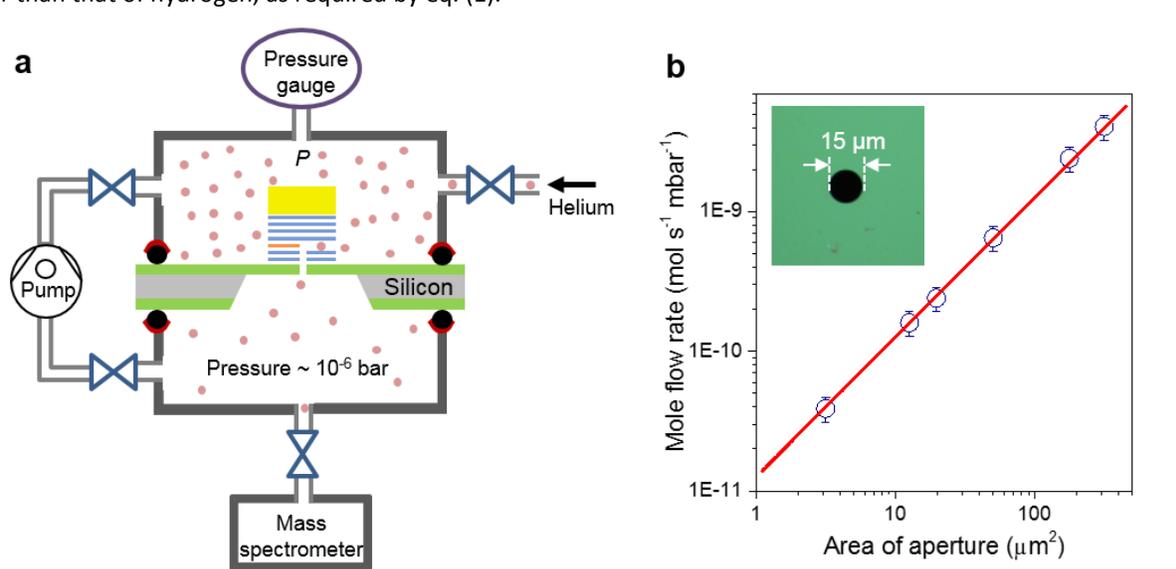

**Supplementary Figure 3| Helium permeation measurements. a,** Schematic of our experimental setup. **b,** He flow through round apertures of various diameters as measured by our He-leak detector (circles). Red line: Expected Knudsen flow through these apertures (no fitting parameters). Inset: Optical image of one of the apertures.

**4. Intrinsic surface roughness of 2D crystals**

To compare flatness of different 2D crystals that served as the slits' walls, we used the density functional theory to calculate near-surface electron density profiles.. The results shown in Supplementary Fig. 4 clearly illustrate that graphene and hBN are atomically more flat than MoS$_2$ as generally expected. For further analysis of the surface roughness, we can employ the criterion of the so-called thermal exclusion surface[35]. It suggests that He atoms effectively 'feel' the surface at a critical density of about 0.03 electron per Å$^3$. Incident atoms with a kinetic energy equivalent to their thermal energy (room temperature) cannot penetrate beyond this isosurface shown by the red curves in Supplementary Fig. 4. The density functional analysis was carried out using the CP2K program[36] and the PBE exchange-correlation functional[37]. The energy cut-off for plane-wave expansions was set at 600 Ry. Gaussian basis sets for double-zeta valence polarized (DZVP) quality[38] and the Goedecker-Teter-Hutter pseudopotentials[39] were used in the calculations. Periodic boundary conditions were applied, and the vacuum region was set to have a thickness of 40 Å. The electron-density contours were analyzed using Multiwfn[40].

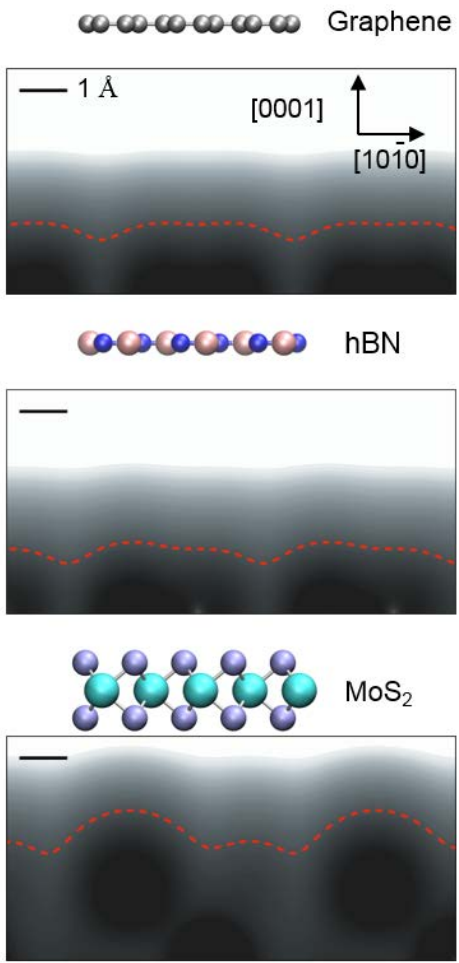

**Supplementary Figure 4| Intrinsic roughness of atomically-flat surfaces.** Electron density profiles at graphite, hBN and MoS$_2$ surfaces. Schematics of their atomic structures are shown on top. The red curves indicate the thermal exclusion surfaces.

**5. Self-cleansing of 2D slits**

Self-cleansing[12,13,25] of interfaces during vdW assembly was intensively studied by many groups over the last five years. As argued in the main text, similar self-cleansing processes should take place not only if two atomically flat surfaces are in direct contact but also at their finite separation. To model this, we take PMMA on graphene as an archetypal example of poorly-mobile adsorbates that are often found on graphene and other surfaces[24-26]. For a PMMA molecule confined inside an angstrom-scale slit, the total free energy has two main contributions: adhesion energy with the two walls and configurational entropy. The former tends to keep PMMA inside whereas the latter increases the total energy if PMMA molecules are flattened and, therefore, pushes them out. For strong confinement (small $h$), the configurational entropy may become dominant and PMMA is squeezed out, which experimentally leads to the formation of contamination bubbles[25]. It is difficult to model the whole squeezing process in MD simulations because of long timescales required for self-cleansing (as witnessed by the necessity of thermal annealing to clean vdW heterostructures) whereas MD analysis allows only simulations lasting typically < 1 msec. Indeed, in our own simulations over such timescales, we could see creep of heavy hydrocarbons but not their complete removal from the simulated nanoslits.

To avoid the timescale problem, we have employed metadynamics algorithms[42] to calculate the potential of mean force

(PMF) which can be regarded as a spatial free energy profile. Also, relatively small PMMA molecules with a molecular weight $M$ of 40 K were simulated for computational reasons because the results are not expected to depend on $M$ (see below). The positions of the center of mass for such PMMA molecules along both relevant directions [parallel ($X$) and perpendicular ($Z$) to slit's axis] were chosen as two variables (Supplementary Data Fig. 5a). Parameters from OPLS forcefield[41] were used to describe interactions among constituent atoms of the PMMA-graphene system, which include bond, angle, dihedral, improper and non-bonded (electrostatics and Lennard-Jones) interactions. Parameters for non-bonded Lennard-Jones interactions were obtained using the Lorentz-Berthelot mixing rules. The PMMA molecule was first placed on top of a graphene sheet, and MD simulations were run in canonical ensembles for 10 ns with a time step of 1 fs, to reach the equilibrium at 298 K. Then, metadynamics simulations were performed for at least 300 ns. All the calculations were carried out using LAMMPS[43].

As an example, Supplementary Fig. 5 shows our results for two graphene capillaries with $N$ = 4 and 12, which correspond to the channel heights $h \approx 13.6$ and 40.8 Å, respectively. For $N$ = 4, the PMMA molecule exhibits a higher energy state inside the capillary than outside (Supplementary Fig. 5b). Therefore, PMMA tends to be squeezed outside or, at least, move to the edge of the entrance, as also illustrated in Fig. 2d of the main text. In contrast, the molecule's free energy is lower inside the taller capillary as shown in Supplementary Fig. 5c and Fig. 2d.

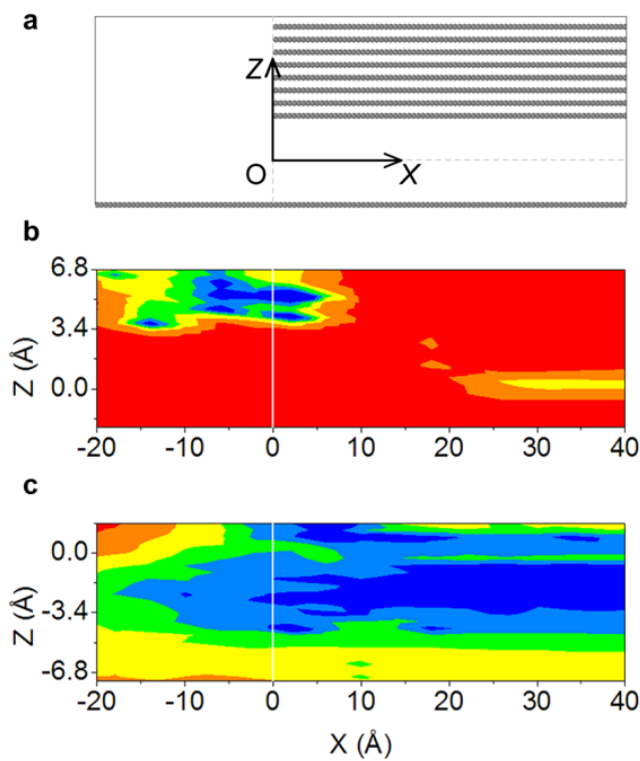

**#Supplementary Figure 5|** MD simulations for a polymer molecule inside angstrom-scale channels. **a**, Sketch of our simulation setup. **b** and **c**, Energy of a PMMA molecule ($M$ = 40 K) for slits with $N$ = 4 and 12, respectively. The axes show the position of the center of mass with respect to the entrance edge. The axes' origin is shown in **a**. Color bars to the right: Relative free energy (PMF).

From another perspective, the height of an adsorbed polymer molecule can also be estimated as $H \approx b/\delta$, where $b$ is the Kuhn length (for PMMA, $b$ = 1.7 nm)[44] and the parameter $\delta$ describes the ratio of the adhesion energy of a monomer to its thermal energy[44]. Our MD simulations for graphene yield $\delta \approx 0.3$ at room temperature and, hence, $H \approx 5.7$ nm. This is the standard estimate, which suggests that the height of adsorbed polymer clumps should not depend on their $M$. To verify the latter assumption, we carried out MD simulations for PMMA on graphene using $M$ from 10 to 200 K. The apparent height $H$ of the clumps was then calculated as an average over 10 ns. The results shown in Supplementary Fig. 6 confirm that $H$ of adsorbed PMMA remains practically constant for $M \geq 40$ K. The obtained $H$ is ~ 4.0 nm, in reasonable agreement with above theory estimate but somewhat smaller. For lighter PMMA molecules, $H$ tends to decrease to 2–3 nm, which is hardly surprising because the statistical chain model is expected to be valid only for long polymers.

The obtained values of $H$ suggest that PMMA contamination could be squeezed out of our slits with $h$ smaller than ~ 4.0 nm ($N \approx 12$) which agrees only qualitatively with the experiment or the simulations in Supplementary Fig. 5. To this end, we note that the apparent height of polymer clumps may be not the best parameter to describe the self-cleansing mechanism. Another possible measure of the height of PMMA molecules is their radius of gyration, $R_g$. Indeed, its perpendicular component $R_{g\perp}$ also provides a sense of height and, therefore, we can define the gyration height as $H_g = 2R_{g\perp}$. For self-cleansing, this parameter seems more meaningful than $H$ because $R_{g\perp}$ refers to the size of a polymer coil[44] and, therefore, implies a direct connection to configuration entropy. We find that $H_g$ for PMMA contamination is about 1.5–2 nm for large $M$ (Supplementary Fig. 6). This value closely matches the results of Supplementary Fig. 5 and our experimental data that indicate the onset of self-cleansing at similar $h$ (see Fig. 2d of the main text). Despite the quantitative agreement, further statistical analysis of self-cleansing would be helpful.

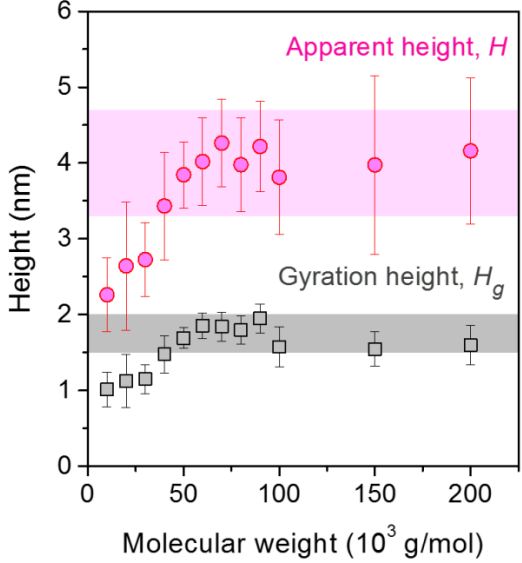

**Supplementary Figure 6| Height of absorbed PMMA.** MD results for the apparent (circles) and gyration (squares) heights of PMMA on graphene. The shaded areas indicate the standard errors using the data for $M \geq$ 40 K.

## 6. Finite size effect for entry of atoms into narrow slits

As discussed in the main text, the Knudsen flow rate through a channel with cross-section $w \times h$ is given by

$$Q = \alpha F w h \quad (S1)$$

where $F = P (m/2\pi RT)^{1/2}$ and $\alpha$ is the transmission coefficient. For 2D channels ($h \ll w < L$) in the Knudsen regime, $\alpha$ can be approximated[6] as

$$\alpha = \frac{h}{L}\left[\frac{\ln\left(h/w + \sqrt{1+(h/w)^2}\right)}{h/w} + \ln\left(\frac{1+\sqrt{1+(h/w)^2}}{h/w}\right) + \frac{1+(h/w)^3 - (1+(h/w)^2)^{3/2}}{3(h/w)^2}\right] \quad (S2)$$

In our case of long and narrow 2D channels, ($h \ll w < L$), $\alpha$ can be approximated as[6]

$$\alpha \approx \frac{16}{3\pi^{3/2}}\frac{h}{L}\ln\left(\frac{4w}{h} + \frac{3h}{4w}\right) \quad \text{where} \quad \frac{16}{3\pi^{3/2}} \approx 0.958 \quad (S3).$$

This can be simplified further as eq. (2) in the main text.

It is instructive to note here that the Smoluchowski correction to eq. (S2) can be modified to make it applicable also in the limit of small $f$, where the factor $(2-f)/f$ diverges, resulting in infinite $Q$. To this end, the flow resistance caused by the entry aperture should be taken into account and, in the first approximation[10], added as a series flow resistance[5,6]. Then the maximum $Q$ is given by eq. (S1) with $\alpha \equiv 1$ and corresponds to fully ballistic transport. In this case, the enhancement coefficient is given by

$$K \equiv Q(\alpha \equiv 1)/Q(\alpha) \approx L/[h \ln(4w/h)] \quad (S4)$$

which yields the dashed magenta curve in Fig. 2a of the main text. This analysis ignores a finite size of transported gas molecules.

Now let us take into account that in our experiments $h$ is comparable with $d$. In the semiclassical description (ignoring a finite de Broglie wavelength), this implies that molecules incident sufficiently away from the channel's center are unable to enter it. For the incidence along the channel axis ($\theta = 0$), the effective channel width $h^*$ then becomes reduced from $h$ to $h - d$ (see Supplementary Fig. 7). For molecules with a non-zero incidence angle $\theta$, the effective width is reduced further to $h^*(\theta) = h - d/\cos\theta$. As an estimate, one may for example choose 45° as an 'average' incidence angle, which yields $h^* = h - \beta d \approx h - \sqrt{2}d$ where the coefficient $\beta$ generally depends on both $d$ and $h$ (see below).

Taking into account the finite size effect, the fully ballistic transport should provide a gas flow

$$Q^* = F w h^* \quad (S5)$$

whereas the enhancement plotted in Fig. 2 by the dashed curve was calculated for $d \equiv 0$. According to eq. (S5), the maximum possible enhancement coefficient $K^*$ which takes into account both ballistic transport ($\alpha \equiv 1$) and the entry effect ($d \neq 0$) is somewhat smaller

$$\kappa^* = Q^*/Q(\alpha) = (h^*/h) K = [1 - (d/h)\beta]K \quad (S6).$$

To estimate $\beta$ more accurately, we consider contributions from all angles $\theta$. Incident atoms effectively 'see' only a projection of the channel opening, which is $h\cos\theta$. If it is smaller than $d$ (that is, if $h^*(\theta) \leq 0$), then those atoms hit one of the aperture's edges. Let us assume in the first approximation that all such scattered atoms are scattered away rather than guided inside the slits. This means that, when integrating, we need to exclude those trajectories that are above the critical angle $\theta_c = \mathrm{acos}(d/h)$ because they do not contribute to $Q^*$. This yields

$$\kappa^* = (K/\pi)\int_{-\pi/2}^{+\pi/2}(1 - d/h\cos\theta)d\theta \quad (S7)$$

where the averaging is carried out over all incident angles $\{-\pi/2, \pi/2\}$ but the function under the integral is zero for $|\theta| \geq |\theta_c|$. We can re-write eqs. (S6&S7) using the average coefficient

$$\langle\beta\rangle = (h/d)(1-2\theta_c/\pi) + (2/\pi)\int_0^{\theta_c} d\theta/\cos\theta \quad (S8).$$

Noting that $\int d\theta/\cos\theta = \ln[(1+\sin\theta)/\cos\theta]$, we obtain

$$\langle\beta\rangle = (h/d)(1-2\theta_c/\pi) + (2/\pi)\ln[(1 +\sin\theta_c)/\cos\theta_c]$$

For the He diameter of 2.6 Å, we find the $\kappa^*$ dependence shown by the solid magenta curve in Fig. 2a of the main text. Our graphene channels with $N = 4$ exhibit the maximal enhancement, and it agrees well with the calculated $\kappa^*$. This again indicates the frictionless helium flow. The finite size effect is expected to be enhanced by the diffraction of de Broglie waves at the entry apertures (for helium, $\lambda_B \approx 0.5$ Å) which should lead to some further reduction in $\kappa^*$ for smallest $N$. We do not expect the quantum correction to be excessively large and, therefore, ignore the effect in our present analysis.

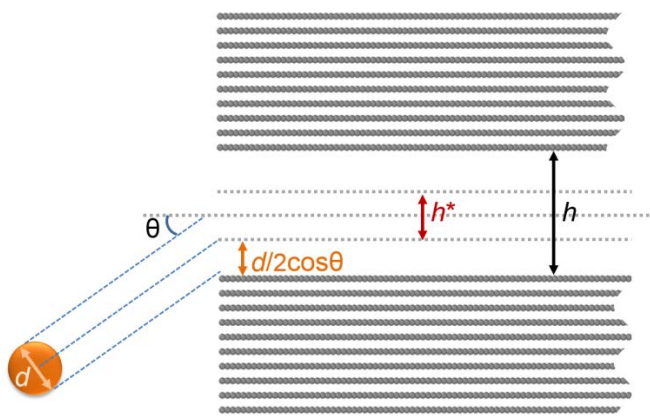

**Supplementary Figure 7| Finite size effect.** If an incident atom of diameter $d$ hits one of the channel edges, it can be reflected. To avoid this, the center trajectory should be $d/2\cos\theta$ away from the edge, effectively reducing the entry aperture to $h^*(\theta) = h - d/\cos\theta$.